\documentclass[10pt]{article}
\usepackage[english]{babel}
\usepackage{amsmath,amsfonts,amssymb,bm}\interdisplaylinepenalty=2500
\ifx\pdfoutput\undefined
  \usepackage{graphicx,color}   
  \usepackage[nativepdf,backref,bookmarks]{hyperref} 
  \usepackage{rotating}
\else
  \usepackage[backref,bookmarks]{hyperref}  
  \usepackage[pdftex]{graphicx,color}     
  \usepackage[pdftex]{rotating}              
\fi
\graphicspath{{figs/}}
\ifx\undefined\degrees\def\degrees{\ensuremath{^{\circ}}}\fi
\ifx\undefined\celsius\def\celsius{\ensuremath{^{\circ}\mathrm{C}}}\fi
\ifx\undefined\unit\def\unit#1{\ensuremath{\mathrm{\,#1}}}\fi
\ifx\undefined\micro\def\micro{\ensuremath{\mu}}\fi
\ifx\undefined\sups\def\sups#1{\ensuremath{^{\mathrm{#1}}}}\fi
\ifx\undefined\subs\def\subs#1{\ensuremath{_{\mathrm{#1}}}}\fi
\ifx\undefined\ohm\def\ohm{\ensuremath{\mathrm{\Omega}}}\fi
\def\req#1{(\protect\ref{#1})}
\begin{document}
\raggedbottom

\title{Advanced bridge (interferometric) phase and amplitude noise measurements}
\author{Enrico Rubiola\thanks{%
    Universit\'e Henri Poincar\'e, Nancy, France,
      \texttt{www.rubiola.org}, e-mail~\texttt{enrico@rubiola.org}}
   \and Vincent Giordano\thanks{%
     Dept.~\textsc{lpmo}, \textsc{femto-st} Besan\c{c}on, France,
     e-mail~\texttt{giordano@lpmo.edu}}}
\date{\small Cite this article as:\\
  E. Rubiola, V. Giordano, ``Advanced interferometric phase and amplitude noise measurements'',  
  \emph{Review of Scientific Instruments} vol.~73 no.~6 pp.~2445--2457, June~2002}

\maketitle

\begin{abstract}
  The measurement of the close-to-the-carrier noise of two-port
  radiofrequency and microwave devices is a relevant issue in time and
  frequency metrology and in some fields of electronics, physics and
  optics.  While phase noise is the main concern, amplitude noise is
  often of interest.  Presently the highest sensitivity is achieved
  with the interferometric method, that consists of the amplification
  and synchronous detection of the noise sidebands after suppressing
  the carrier by vector subtraction of an equal signal.  A substantial
  progress in understanding the flicker noise mechanism of the
  interferometer results in new schemes that improve by 20--30 dB the
  sensitivity at low Fourier frequencies.  These schemes, based on two
  or three nested interferometers and vector detection of noise, also
  feature closed-loop carrier suppression control, simplified
  calibration, and intrinsically high immunity to mechanical
  vibrations.
  
  The paper provides the complete theory and detailed design criteria,
  and reports on the implementation of a prototype working at the
  carrier frequency of 100 MHz.  In real-time measurements, a
  background noise of $-175$ to $-180$ dB \unit{dBrad^2/Hz} has been
  obtained at $f=1$ Hz off the carrier; the white noise floor is
  limited by the thermal energy $k_BT_0$ referred to the carrier power
  $P_0$ and by the noise figure of an amplifier.  Exploiting
  correlation and averaging in similar conditions, the sensitivity
  exceeds $-185$ \unit{dBrad^2/Hz} at $f=1$ Hz; the white noise floor
  is limited by thermal uniformity rather than by the absolute
  temperature.  A residual noise of $-203$ \unit{dBrad^2/Hz} at
  $f=250$ Hz off the carrier has been obtained, while the ultimate
  noise floor is still limited by the averaging capability of the
  correlator.  This is equivalent to a $S/N$ ratio of
  $2{\times}10^{20}$ with a frequency spacing of $2.5{\times}10^{-6}$.
  All these results have been obtained in a relatively unclean
  electromagnetic environment, and without using a shielded chamber.
  Implememtation and experiments at that sensitivity level require
  skill and tricks, for which a great effort is spent in the paper.
  
  Applications include the measurement of the properties of materials
  and the observation of weak flicker-type physical phenomena, out of
  reach for other instruments.  As an example, we measured the flicker
  noise of a by-step attenuator ($-171$ \unit{dB[rad^2]/Hz} at $f=1$
  Hz) and of the ferrite noise of a reactive power divider ($-173.7$
  \unit{dB[rad^2]/Hz} at $f=1$ Hz) without need of correlation.  In
  addition, the real-time measurements can be exploited for the
  dynamical noise correction of ultrastable oscillators.
\end{abstract}

\setcounter{tocdepth}{2}        
\tableofcontents

\section{Introduction}

The output signal of a two-port device under test (DUT) driven by a
sinusoidal signal of frequency $\nu_0$ can be represented as
\begin{equation}
x(t)=V_0[1+\alpha(t)]\cos[2\pi\nu_0t+\phi(t)]
\label{eqn:polar:dut-signal}
\end{equation}
where $\phi(t)$ and $\alpha(t)$ are the random phase and the random
normalized amplitude fluctuation of the DUT, respectively.
Close-to-the-carrier noise is usually described in term of $S_\phi(f)$
and $S_\alpha(f)$, namely the phase spectrum density (PSD) of
$\phi(t)$ and $\alpha(t)$ as a function of the Fourier frequency $f$.
$\phi(t)$ and $\alpha(t)$ originate from both additive and parametric
noise contributions, the latter of which is of great interest because
it brings up the signature of some physical phenomena.  True random
noise is locally flat ($f^0$) around $\nu_0$.  Conversely, parametric
noise contains flicker ($f^{-1}$) noise and eventually higher slope
noise processes as $f$ approaches zero.

The instrument of the interferometric type, derived from early
works~\cite{sann68mtt,labaar82microw}, show the highest sensitivity; new
applications for them have been reported~\cite{ivanov98mtt}. Two recent
papers provide insight and new design rules for general and real-time
measurements~\cite{rubiola99rsi} and give the full explanation of the white
noise limit in correlation-and-averaging measurements~\cite{rubiola00rsi}.
The residual flicker of these instruments turned out to be of $-150$
\unit{dBrad^2/Hz} at 1 Hz off the carrier for the real-time version,
and $-155$ \unit{dBrad^2/Hz} correlating two interferometers.

The scientific motivations for further progress have not changed in
the past few years.  Nonetheless, we whish to stress the importance of
close-to-the-carrier noise for ultrastable oscillators.  First,
oscillators, inherently, turns phase noise into frequency
noise~\cite{leeson66pieee}, which makes the phase diverge in the long run.
Then, amplitude noise affects the frequency through the the resonator
sensitivity to power, as it occurs with quartz
crystals~\cite{gagnepain75} and microwave whispering gallery mode
resonators~\cite{chang97prl}.  Finally, the knowledge of the
instantaneous value of $\phi(t)$ and $\alpha(t)$ in real time enables
additional applications, such as the dynamical noise compensation of a
device, for which the statistical knowledge is insufficient.

This paper is the continuation of two previous ones~\cite{rubiola99rsi,rubiola00rsi}
in this field.  After them, several elements of progress have been
introduced, the main of which are: 1) the flicker noise mechanism has
been understood, 2) the carrier suppression adjustment has been split
into coarse and fine, 3) elementary algebra has been introduced to
process signals as complex vectors, 4) the carrier suppression has
been treated as a complex virtual ground.  This results in new design
rules and in a completely new scheme that exhibit lower residual
flicker and increased immunity to mechanical vibrations.  Calibration
is simplified by moving some issues from radiofrequency hardware to
the detector output.  Finally, the carrier suppression is controlled
in closed-loop, which is a relevant point for at least two reasons.
Firstly, the interferometer drifts, making the continuous operation of
the instrument be impossible in the long run.  Then, the residual
carrier affects the instrument sensitivity through different
mechanisms, and a sufficient suppression can only be obtained in
closed-loop conditions.

\section{The interferometer revisited}
%
\begin{figure}[t]
\centering\includegraphics[scale=1]{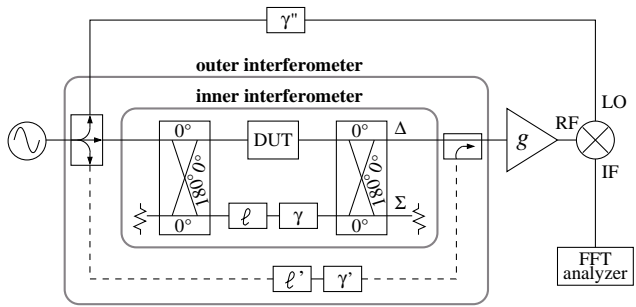}
\caption{Basic scheme of the interferometric phase noise detection.}
\label{fig:basic-sch}
\end{figure}
A digression about the interferometric noise measurement instruments
is needed prior to develop the complete scheme.  The starting point is
the scheme of Fig.~\ref{fig:basic-sch}, which includes the major ideas
of our previous papers~\cite{rubiola99rsi,rubiola00ell,rubiola01fcs,rubiola02uffc}, plus
several unpublished ideas.

The key idea of the interferometric method is that phase noise---as
well as amplitude noise---resides entirely in the sidebands, and that
several advantages arise from removing the carrier signal.  Thus,
matching the attenuator $\ell$ and the phase shifter $\gamma$ to the
device under test (DUT), the carrier is suppressed at the output
$\Delta$ of the right hybrid.  The DUT noise sidebands, not affected
by the above equilibrium condition, are amplified and down converted
to baseband by synchronous detection.  Properly setting the phase
$\gamma''$, the machine detects the instantaneous value of $\phi(t)$
or $\alpha(t)$, or the desired combination.  Basically, the
interferometer is an impedance-matched null bridge; the detector can
be regarded as a part of lock-in amplifier~\cite{meade:lock-in} or of a
phase-coherent receiver~\cite{lindsey:telecomm,viterbi:communication}.

The instruments of the first generation~\cite{rubiola99rsi} make use of
continuously variable attenuators and phase shifters as $\ell$ and
$\gamma$; the dotted path, with $\ell'$ and $\gamma'$, is absent.  A
carrier rejection of some 70--80 dB can be obtained, limited by the
resolution and by the stability of $\ell$ and $\gamma$; the adjustment
requires patience and some skill.  Experimenting on interferometers at
10 MHz, 100 MHz and 7--10 GHz, the achievable carrier rejection turned
out to be of the same order of magnitude.

The second-generation instruments~\cite{rubiola00rsi} make use of correlation
and averaging to reduce the residual noise.  This approach results in
outstanding white noise, not limited by the thermal energy $k_BT_0$
referred to the carrier power $P_0$; $k_B=1.38{\times}10^{-23}$ J/K is
the Boltzmann constant, and $T_0\simeq295$ K is the room temperature.

The third-generation instruments~\cite{rubiola00ell,rubiola01fcs,rubiola02uffc} show
improved sensitivity at low Fourier frequencies, where flicker is
dominant.  This feature is provided by $\ell'$ and $\gamma'$ and the
dashed path.

Improving the low-frequency sensitivity relies upon the following
issues, which updates the previous design rules~\cite{rubiola99rsi}.

\paragraph{Amplifier Noise.}  
The basic phenomenon responsable for the close-to-the-carrier flicker
noise of amplifiers is the up conversion of the near-dc flickering of
the bias current, due to carrier-induced nonlinearity.  Interesting
analyses are available for bipolar
transistors~\cite{walls97uffc,kuleshov97fcs}, yet this phenomenon is general.
In fact, if the carrier power is reduced to zero, the noise spectrum
at the output of the amplifier is of the white type, and no $1/f$
noise can be present around the carrier frequency $\nu_0$.  The
assumption is needed that the DUT noise sidebands be insufficient to
push the amplifier out of linearity, which is certainly true with low
noise DUTs.

After Friis~\cite{friis44ire}, it is a common practice to calculate the
white noise of a system by adding up the noise contribution of each
stage divided by the gain of all the preceding ones, for the first
stage is the major contributor.  But flicker noise behaves quite
differently.  Let us consider the design of a low-noise small-signal
amplifier based on off-the-shelf parts.  Almost unavoidably, the
scheme ends up to be a chain of modules based on the same technology,
with the same input and output impedance, and with the same supply
voltage.  Therefore, the bias current and the nonlinear coefficients
are expected to be of the same order.  Consequently, flicker noise
tends to originate from the output stage, where the carrier is
stronger, rather than from the first stage.

\paragraph{Interferometer Stability.} 
Common sense suggests that the flicker noise of the interferometer is
due the mechanical instability of the variable elements $\ell$ and
$\gamma$ and of their contacts, rather than to the instability of the
semirigid cables, connectors, couplers, etc.  By-step attenuators and
phase shifters are more stable than their continuously adjustable
counterparts because the surface on which imperfect contacts fluctuate
is nearly equipotential; that contacts flicker is a so well
estabilished fact that Shockley~\cite{shockley:semiconductors} uses
the term `contact-noise' for the $1/f$ noise.  Even higher stability
is expected from fixed-value devices, provided the carrier suppression
be obtained.

\paragraph{Resolution of the Carrier Suppression Circuit.}
An amplitude error of $0.05$ dB, which is half of the minimum step for
off-the-shelf attenuators, results in a carrier rejection of 45 dB;
accounting for a similar contribution of the phase shifter, the
carrier rejection is of 42 dB in the worst case.  This is actually
insufficient to prevent the amplifier from flicker.

\paragraph{Rejection of the Oscillator Noise.}
The difference in group delay between the two arms of the
interferometer acts as a discriminator, for it causes a fraction of
the oscillator phase noise to be taken in; the effect of this can be
negligible if the DUT delay is small.  Conversely, the rejection of
the oscillator amplitude noise relies upon the carrier rejection at
the amplifier input.

\paragraph{Dual Carrier Suppression.}
A high carrier rejection is obtained with two nested interferometers.
The inner one provides a high stability coarse adjustment of the phase
and amplitude condition, while the outer one provides the fine
adjustment needed to interpolate between steps.  Owing to the small
weight of the interpolating signal, as compared to the main one,
higher noise can be tolerated.  An additional advantage of the nested
interferometer scheme is the increased immunity to mechanical
vibrations that results from having removed the continuously
adjustable elements from the critical path.

In an even more complex version of the nested interferometer, the
amplifier is split in two stages, and the correction signal is
injected in between~\cite{rubiola00ell}.  The game consists of the
gain-linearity tradeoff, so that the residual carrier due to the
by-step adjustment is insufficient to push the amplifier out of
linearity.  A residual flicker $S_\phi(1\:\unit{Hz})=-160$
\unit{dBrad^2/Hz} has been obtained.  Experimenting on both the
configuration similar residual noise has been
obtained~\cite{rubiola01fcs,rubiola02uffc}.

\paragraph{Hybrid couplers and power splitters.}
Basically, a reactive power splitter is a hybrid coupler internally
terminated at one port (when the termination has not a relevant role,
we let it implicit using a simpler symbol).  The choice between
Wilkinson power splitters, 180\degrees\ hybrids, and 90\degrees\ 
hybrids is just a technical problem.  The signal available at the
$\Sigma$ port of the interferometer should not be used to pump the
mixer, unless saving some amount of power is vital; otherwise the
finite $\Sigma{-}\Delta$ isolation makes the adjustment of the carrier
suppression interact with the calibration of $\gamma''$.

\section{The I-Q Controlled Interferometer}\label{sec:iq-interferometer}
%
\begin{sidewaysfigure}
\centering\includegraphics[scale=1]{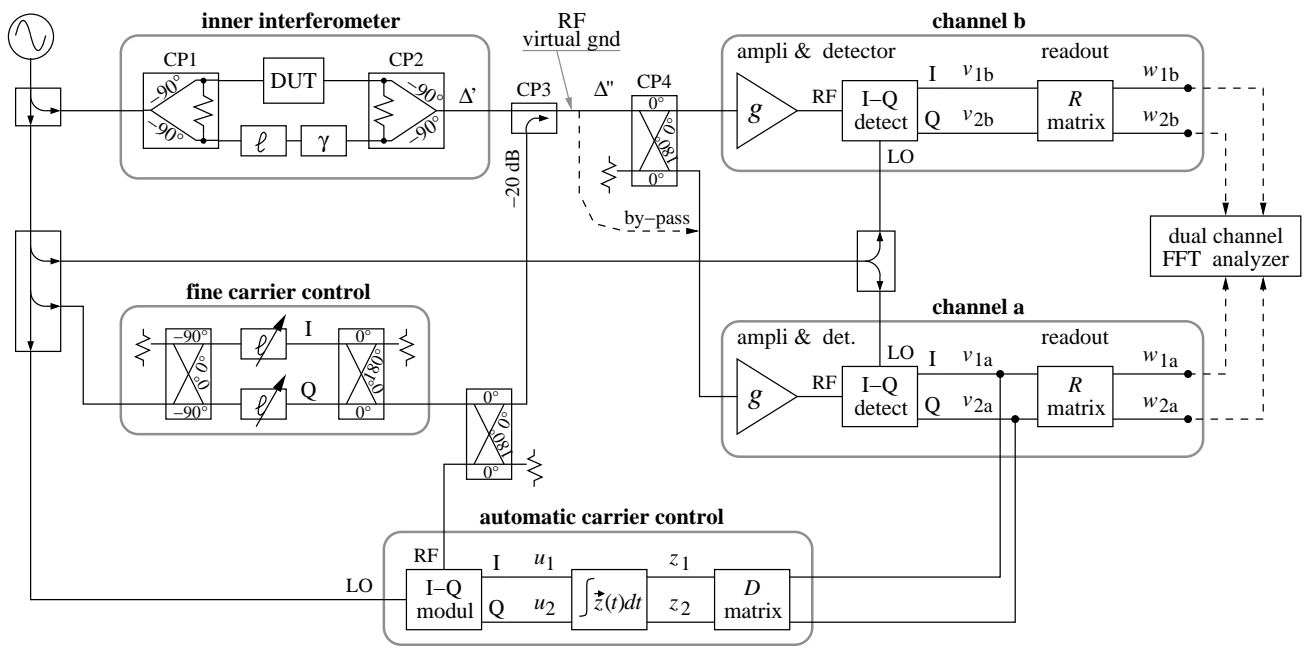}
\caption{Scheme of the proposed instrument.}
\label{fig:scheme}
\end{sidewaysfigure}
\begin{figure}[t]
\centering\includegraphics[scale=1]{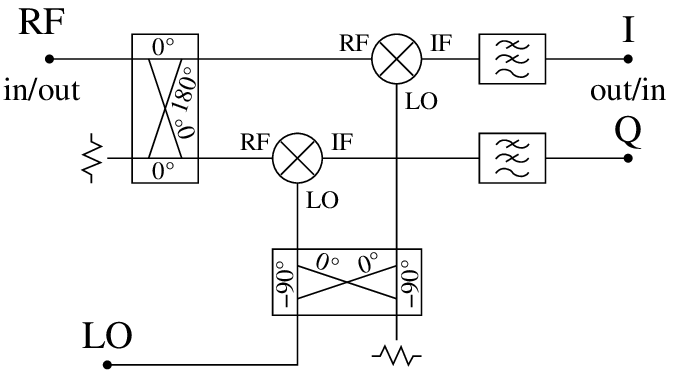}
\caption{Scheme of the I-Q detector-modulator.}
\label{fig:iq}
\end{figure}
Figure~\ref{fig:scheme} shows the scheme of the proposed instrument,
and Fig.~\ref{fig:iq} details the I-Q modulator-detector.  In order to
analyze the detection of the DUT noise we assume that all the
components but the mixers are ideal and lossless, and we also neglect
the intrinsic loss of the 20 dB coupler; the corrections will be
introduced later.  The mixers show a single side band (SSB) loss
$\ell_m$, which accounts for intrinsic and dissipative losses; this is
consistent with most data sheets of actual components.

Basically, the instrument works as a synchronous receiver that detects
the DUT noise sidebands.  Let $N\subs{rf}$ be the PSD of the DUT noise
around the carrier; the dimension of $N\subs{rf}$ is W/Hz, thus
dBm/Hz.  By inspection on the scheme, the noise at the mixer input is
$gN\subs{rf}/8\ell_m$ , thus $gN\subs{rf}/4\ell_m$ at each output of
the I-Q detectors; this occurs because the power of the upper and
lower sidebands is added in the detection process.  The PSD of the
output voltage, either $v_1$ or $v_2$, is
$S_v(f)=gR_0N\subs{rf}/4\ell_m$, where $R_0$ is the mixer output
impedance.  Hence the dual side band (DSB) gain (or noise gain), which
is defined as
\begin{equation}
k\subs{dsb}=\sqrt{\frac{S_v}{N\subs{rf}}}\;\;,
\label{eqn:def-kdsb}
\end{equation}
is
\begin{equation}
k\subs{dsb}=\sqrt{\frac{gR_0}{4\ell_m}}\;\;.
\label{eqn:kdsb}
\end{equation}
$k\subs{dsb}$ it is a constant of the machine, and it is independent
of the DUT power $P_0$.  Yet, in the calibration process it is
convenient to measure the SSB gain $k\subs{ssb}=k\subs{dsb}/\sqrt{2}$.

Getting closer into detail, the signal at the output of an actual DUT
can be rewritten as
\begin{equation}
x(t)=V_0\cos(2\pi\nu_0t)+n_1(t)\cos(2\pi\nu_0t)-n_2(t)\sin(2\pi\nu_0t)\;\;,
\label{eqn:dut-signal}
\end{equation}
which is equivalent to \req{eqn:polar:dut-signal} in low noise
conditions; although \req{eqn:dut-signal} is used to describe the
close-to-the-carrier noise, it is not a narrowband
representation~\cite{davenport:noise}. The polar representation
\req{eqn:polar:dut-signal} is related to the Cartesian one
\req{eqn:dut-signal} by
\begin{eqnarray}
\displaystyle
\alpha(t)&=& \frac{n_1(t)}{V_0} 
         \label{eqn:alpha:vs:n}\\
\displaystyle
\phi(t)&=&  \frac{n_2(t)}{V_0}
         \label{eqn:phi:vs:n}\;\;
\end{eqnarray}
After removing the carrier from  
\req{eqn:dut-signal}, the signals at the detector output are
\begin{eqnarray}
\displaystyle
v_1(t)&=&\sqrt{\frac{g}{8\ell_m}}\;
\left[n_1(t)\cos\psi-n_2(t)\sin\psi\right] \label{eqn:mixer:i:out}\\
\displaystyle
v_2(t)&=&\sqrt{\frac{g}{8\ell_m}}\;
         \left[n_1(t)\sin\psi+n_2(t)\cos\psi\right]
         \label{eqn:mixer:q:out}\;\;,
\end{eqnarray}
where $\psi$ is the arbitrary phase that derives from the phase lag
difference between the input and the pump signal of the I-Q detector.
Setting $\psi=0$, channel 1 detects the phase noise only and channel 2
detects the amplitude noise only, thus $v_1(t)=k_\alpha\alpha(t)$ and
$v_2(t)=k_\phi\phi(t)$, where
\begin{eqnarray}
\displaystyle
k_\alpha & = & \sqrt{P_0}\,k\subs{dsb}  \label{eqn:k-alpha}\\
\displaystyle
k_\phi   & = & \sqrt{P_0}\,k\subs{dsb}  \label{eqn:k-phi}
\end{eqnarray}

In the earlier insrtruments we set $\psi$ acting on a phase shifter in
series to the mixer pump ($\gamma''$ in Fig.~\ref{fig:basic-sch}),
which is uncomfortable.  Now we prefer to let $\psi$ arbitrary and to
process the output signals, as described in the next Section.

In the absence of the DUT, the equivalent noise at the amplifier input
is $Fk_BT_0$, where $F$ is the amplifier noise figure. Thus the noise
PSD at the mixer output is
\begin{equation}
S_{v0}=\frac{gFk_BT_0R_0}{\ell_m} \;\;.  \label{eqn:s-v0}
\end{equation}
$S_{v0}$ is a constant of the instrument, independent of $P_0$.
Dividing \req{eqn:s-v0} by $k_\phi$, or by $k_\alpha$, we get the
phase and amplitude noise floor
\begin{equation}
S_{\phi0}=S_{\alpha0}=\frac{4Fk_BT_0}{P_0}\;\;. 
\label{eqn:s-phi-0}
\end{equation}

If only one of the two radiofrequency channels is used, and the
splitter in between is bypassed, the DSB gain \req{eqn:kdsb} becomes
$k\subs{dsb}=\sqrt{gR_0/2\ell_m}$; as $k\subs{dsb}$ is multiplied by
$\sqrt{2}$, also $k_\alpha$ and $k_\phi$ are; thus
$k_\alpha=k_\phi=\sqrt{2P_0}k\subs{dsb}$.  Therefore, as $S_{v0}$ is a
constant, the phase and amplitude noise floor become
$S_{\phi0}=S_{\alpha0}=2Fk_BT_0/P_0$.

Finally, the effect of all the dissipative losses in the DUT-mixer
path, plus the insertion loss of the 20 dB coupler CP3 (this accounts
for dissipative and intrinsic losses, as in the data sheet of actual
components), is to decrease $k\subs{ssb}$, thus $k_\alpha$ and
$k_\phi$.  The effect of all the dissipative losses in the
DUT-amplifier path, plus the insertion loss of CP3, is to increase
$S_{\phi0}$ and $S_{\alpha0}$, letting $S_{v0}$ unaffected.

\section{Readout}\label{sec:readout}
%
\begin{figure}[t]
\centering\includegraphics[scale=1]{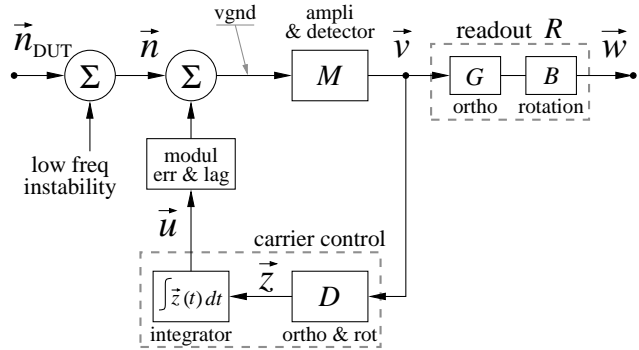}
\caption{Information flow of the instrument.  
  For clarity, some amplitude coefficients are omitted and only one
  channel is shown.}
\label{fig:info-flow}
\end{figure}
Figure \ref{fig:info-flow} shows the information flow through the
instrument.  This scheme is equivalent to that of
Fig.~\ref{fig:scheme}, but the radiofrequency circuits are hidden, for
all the radiofrequency signals are replaced with their baseband
representation in terms of Fresnel vector.  As an example, the noise
of \req{eqn:dut-signal} takes the form $\vec{n}(t)=[n_1(t),n_2(t)]^T$,
where $T$ stands for transposed.  It is assumed in this Section that
the carrier control works properly without interferring with the DUT
noise, hence we account for the DUT noise $\vec{n}(t)$ only, omitting
the subscript \textsc{dut}.

The signal $\vec{v}(t)=[v_1(t),v_2(t)]^T$ at the output of the
radiofrequency section is transformed into the desired signal
$\vec{w}(t)=[w_1(t),w_2(t)]^T=k\,[\alpha(t),\phi(t)]^T$ through the
transformation $\vec{w}(t)=R\vec{v}(t)$, where $R=[r_{ij}]$ is the
$2{\times}2$ readout matrix.  Equations \req{eqn:mixer:i:out} and
\req{eqn:mixer:q:out} turn into the relationship
$\vec{v}(t)=M\vec{n}(t)$, i.e.\ 
\begin{equation}
\left[\begin{array}{c}v_1(t)\\v_2(t)\end{array}\right] =
\left[\begin{array}{cc}m_{11}&m_{12}\\m_{21}&m_{22}\end{array}\right]
\left[\begin{array}{c}n_1(t)\\n_2(t)\end{array}\right]  \;\;,
\label{eqn:define-m-matrix}
\end{equation}
which also accounts for the gain error and asymmetry of the two
channels and for the quadrature error of the I-Q detector.  Thus, the
matrix $R$ provides the frame transformation by which
\begin{equation}
\vec{w}(t)=RM\:\vec{n}(t) \;\;.  \label{eqn:matrix-r}
\end{equation}
Of course, the appropriate $R$ is the solution of $RM=aI$, where $I$
is the identity matrix and $a$ a constant, so that
$\vec{w}(t)=a\vec{n}(t)$.

The direct measurement of $M$ relies upon the availability of two
reference vectors that form a base for $\vec{n}$, the simplest of
which is $\vec{v}\,'=[1, 0]^T$ and $\vec{v}\,''=[0,1]^T$.  This can be
done by means of a reference AM-PM modulator at the DUT output, or by
means of a reference I-Q modulator that introduces a reference signal
in the DUT path; unfortunately, both these solutions yield impractical
calibration aspects.  Therefore, we split the problem in two tasks,
which is accomplished by letting $R=BG$; the matrix $G=[g_{ij}]$ makes
the detection axes orthogonal and symmetrical, while $B=[b_{ij}]$
rotates the frame.

We first find $G$ with the well known Gram-Schmidt
process~\cite{horn:matrix.1}, replacing the DUT output signal with a
pure sinusoid $V_s\cos(2\pi\nu_st)$.  To do so, the DUT, the variable
attenuator and the variable phase shifter are temporarily removed from
the inner interferometer, and all the unused ports are terminated.
The frequency $\nu_s$ is set just above $\nu_0$, so that the detected
signal be a tone at the frequency $f_s=\nu_s-\nu_0$ of a few kHz.  The
driving signal is equivalent to the vector
\begin{equation}
\left[\begin{array}{c}n_1(t)\\n_2(t)\end{array}\right] = 
\left[\begin{array}{c}V_s\cos(2\pi f_st)\\
                      V_s\sin(2\pi f_st)
\end{array}\right] \;\;.
\label{eqn:sideband-equiv-vec}
\end{equation}
Setting $R=I$, thus $B=I$ and $G=I$, we measure the output signals
$W_1$ and $W_2$ by means of the dual-channel FFT analyzer; $W_1$ and
$W_2$ are the rms values of the corresponding signals $w_1(t)$ and
$w_2(t)$.  The result consists of the squared modules $|W_1|^2$ and
$|W_2|^2$ and the cross signal $W_{12}=|W_1||W_2|\cos\theta +
j|W_1||W_2|\sin\theta$, where $\theta$ is the angle formed by the two
signals.  Then, setting\footnote{The original article (RSI~\textbf{73}~6) contains some
typesetting errors in Eq.~(16) and (17).  The Equations \req{eqn:g-12} and \req{eqn:g-solution} reported here are correct.}
\begin{equation}
g_{21}=-\frac{\Re\{W_{12}\}}{|W_1|^2} 
\label{eqn:g-12}
\end{equation}
the two detection channels are made orthogonal, but still
asymmetrical.  To correct this, we measure $W_1$ and $W_2$ in this new
condition, and we set
\begin{equation}
g_{21}=\left|\frac{W_1}{W_2}\right| g\sups{(p)}_{21}
\qquad\mathrm{and}\qquad
g_{22}=\left|\frac{W_1}{W_2}\right| g\sups{(p)}_{22} 
\label{eqn:g-solution}
\end{equation}
where the superscript (p) stands for the previous value.  Now the two
channels are orthogonal and of equal gain. 

Turning equations into laboratory practice, this is the right place
for the measurement of $k\subs{ssb}$.  Letting $P_s$ the power of the
sideband at the DUT output, we get
\begin{equation}
k\subs{ssb}=\frac{W}{\sqrt{P_s}}\;\;,
\label{eqn:k-ssb-meas} 
\end{equation}
where the subscript of $W$ is omitted since now it holds $W_1=W_2$.

In order to complete the task we still have to calculate the rotation
matrix $B$, for we need a reference to set the origin of angles.
After reassembling the inner interferometer, we insert as the DUT a
phase modulator driven by a reference sinusoid and we measure the
output signals $W_1$ and $W_2$; the method works in the same way with
pseudo-random noise, which is preferable because of the additional
diagnostic power.  $R$ is temporarily let equal to $G$, thus $B=I$; in
this conditions we measure $W_1$ and $W_2$
\begin{eqnarray}
W_1 & = & V_s\cos\beta      \label{eqn:w-1}\\
W_2 & = & V_s\sin\beta      \label{eqn:w-2}\;\;,
\end{eqnarray}
from which we calculate the frame rotation $\beta$.  Finally, a
rotation of $-\beta$ is needed, performed by
\begin{equation}
B = \left[\begin{array}{cc}
      \cos\beta  & -\sin\beta\\
      \sin\beta  & \cos\beta
    \end{array}\right] \;\;.
\label{eqn:matrix-b}
\end{equation}

Due to the hardware, it might be necessary to scale $R$ up or down
during the process.  In our implementation, for instance, there is the
constraint $|r_{ij}|\le 1$, $\forall ij$.

\section{Automatic Carrier Suppression}\label{sec:control}

The carrier suppression circuit of Fig.~\ref{fig:scheme} works
entierly in Cartesian coordinates.  This is obtained by means of an
I-Q modulator that controls the amplitude of two orthogonal phases of
a signal added at the amplifier input, which nulls separately the real
and imaginary part of the residual carrier.  This method, which is
somewhat similar to the vector voltage-to-current-ratio measurement
scheme used in a low-frequency impedance analyzer~\cite{agilent:hp4192a}, can
be regarded as complex virtual ground.

With reference to Fig.~\ref{fig:info-flow}, the system to be
controlled transforms the input signal $\vec{u}(t)=[u_1(t), u_2(t)]^T$
into $\vec{v}(t)=[v_1(1), v_2(t)]^T$ through
\begin{equation}
\vec{v}(t) = A\:\vec{u}(t) \;\;.
\end{equation}
The $2{\times}2$ matrix $A$ models the gain and the rotation that
result from all the phase lags in the circuit; $A$ also accounts for
the gain asymmetry and for the quadrature error of the I-Q modulator
and detector.  Introducing the $2{\times}2$ diagonalization matrix
$D$, we get
\begin{equation}
\vec{z} = DA\:\vec{u} \;\;, 
\end{equation}
The appropriate $D$ is the solution of $DA=cI$, where $c$ is a
constant, thus
\begin{equation}
D = \frac{c}{\mathrm{det}\,A}
    \left[\begin{array}{cc}
        a_{22} & -a_{12} \\
       -a_{21} &  a_{11} 
    \end{array}\right] \;\;.
\label{eqn:d:solution}
\end{equation}
Therefore, the two-dimensional control is split in two independent
control loops.  This is a relevant point because interaction could
result in additional noise or in a chaotic behavior.  Actually, the
quadrature error of the I-Q devices is relatively small, hence
\req{eqn:d:solution} is free from the error enhancement phenomena
typical of ill-conditioned problems.

$A$ consists of the four voltage gains
\begin{equation}
a_{ij} = \frac{v_i}{u_j} \;\;,
\end{equation}
which is easily measured with the transfer-function capability of the
FFT spectrum analyzer.  Pseudo-random noise is preferable to a simple
tone because of its diagnostic power.

A simple integral
\begin{math}
\vec{u}(t)=-a_0\int D\:\vec{v}(t)\;\mathrm{d}t
\end{math}
is sufficient to control the carrier without risk of oscillation or
instability.  This occurs because there are no fast variations to
track and because, even with simple electronics, the poles of $DA$ end
up to be at frequencies sufficiently high not to interact with the
control.

In the normal operating mode, the cutoff frequency $f_0=a_0/2\pi$ of
the control loop must be lower than the lowest Fourier frequency of
interest, and a margin of at least one decade is recommended.  An
alternate mode is possible~\cite{audoin81,chronos:frequency}, in which
the control is tight, and the DUT noise is derived from the error
signal. We experimented on the normal mode only.

It should be stressed that the phase and amplitude of the DUT output
signal do not appear---explicitely or implied---in the equations of
the control loop.  As a relevant consequence, no change to the control
parameters $A$ and $a_0$ is necessary after the first calibration,
when the instrument is built.

The automatic carrier suppression of this machine turns into a
difficult problem if not approached correctly, for we give additional
references.  A fully polar control based on a phase and amplitude
detector and on a phase and amplitude modulator, similar to that used
to extend the dynamic range of spectrum analyzers by removing a
`dazzling' carrier~\cite{horn69fcs}, suffers from the basic difficulty
that the phase becomes undefined as the residual signal approaches
zero. The mixed polar-Cartesian control, based on a phase and
amplitude modulator as the actuator and on a mixer pair as the
detector, is simpler than our scheme; it has been successfully used to
stabilize a microwave oscillator~\cite{ivanov98mtt}.  Yet, the mixed
control is incompatible with the nested interferometer scheme because
the residual carrier, made small by the inner interferometer, spans
over a wide range of relative amplitude, for the loop gain of the
phase channel is unpredictable and can also change sign.  In the field
of telecommunications, the polar-loop control was proposed as a means
to linearize the power amplifier in SSB
transmitters~\cite{petrowich79ell}, but the advantages of a fully
Cartesian-frame control were soon
recognized~\cite{kennington:hi-lin-ampli}.

\section{Correlation Techniques}\label{sec:correlation}

The cross power spectrum density $S_{ab}(f)$ is
\begin{equation}
S_{ab}(f) = \mathcal{F}\left\{\mathcal{R}_{ab}(\tau)\right\}
          = \int_\infty\mathcal{R}_{ab}(\tau)\exp(-2\pi f\tau)\;d\tau
\end{equation}
were $\mathcal{F}\{.\}$ is the Fourier transform operator, and
$\mathcal{R}_{ab}(\tau)$ is the cross correlation function
\begin{equation}
\mathcal{R}_{ab}(\tau) = 
\lim_{\theta\rightarrow\infty}
\frac{1}{\theta}\int_\theta a(t)\:b^\ast(t-\tau)\;dt \;\;;
\end{equation}
as we measure real signals, the complex conjugate symbol `$\ast$' can
be omitted.  $S_{ab}(f)$ is related to the Fourier transform $A(f)$
and $B(f)$ of the individual signals by
\begin{equation}
S_{ab}(f)=A(f)\:B^\ast(f)\;\;,
\end{equation}
which is exploited by dynamic signal analyzers; the Fourier transform
is replaced with the FFT of $a(t)$ and $b(t)$ sampled simultaneously,
and the spectrum is averaged on a convenient number $m$ of
acquisitions; the rms uncertainty is $\sigma=|A|\,|B|/\sqrt{2m}$.
Both averaging and Fourier transform are linear operators, for $a(t)$
and $b(t)$ can be divided into correlated and uncorrelated part, that
are treated separately.  With the uncorrelated part, $S_{ab}(f)$
approaches zero as $1/\sqrt{2m}$, limited by $\sigma$.  This is
exploited to extend the sensitivity beyond the thermal energy limit
$k_BT_0$.

\subsection{Parallel Detection}\label{sec:parallel-correlation}

In the normal correlation mode the matrices $R$ are set for the two
channels to detect the same signal, thus $w_{1}(t) \propto n_1(t)$ and
$w_{2}(t) \propto n_2(t)$ if only the DUT noise is present.

\begin{figure}[t]
\centering\includegraphics[scale=1]{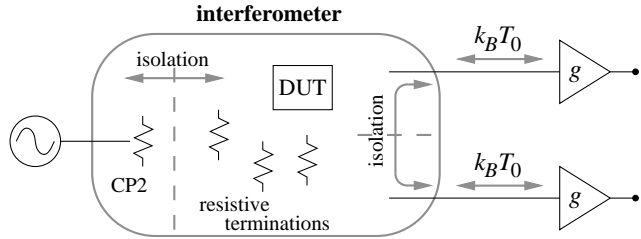}
\caption{Thermal-noise model of the interferometer.}
\label{fig:noise-model}
\end{figure}

Let us analyze the instrument in the presence of thermal noise only,
coming from the DUT and from the resistive terminations, under the
assumption that the temperature is homogeneous.  As there are several
resistive terminations, the complete signal analysis~\cite{rubiola00rsi} is
unnecessarily complicated, thus we derive the behavior from physical
insight.  The machine can be modeled as in Fig.~\ref{fig:noise-model}.
All the oscillator power goes to one termination---or to a set of
terminations---isolated from the rest of the circuit; the amplifier
inputs are isolated from one another and from the oscillator.  In
thermal equilibrium, a power per unit of bandwidth $k_BT_0$ is
exchanged between the input of each amplifier and the instrument core.
The two signals flowing into the amplifiers must be uncorrelated,
otherwise the second principle of thermodynamics would be violated.
Consequently, the thermal noise yields a zero output.

As a consequence of linearity, the non-thermal noise of the DUT is
detected, and the instrument gain ($k\subs{ssb}$, or $k_\alpha$ and
$k_\phi$), as derived in Section~\ref{sec:iq-interferometer} applies.
The instrument measures extra noise (we avoid the term ``excess
noise'' because it tend to be as a synonymous of flicker, which would
be restrictive), even if it is lower than the thermal energy in the
same way as the double interferometer~\cite{rubiola00rsi} does.  This is the
same idea of the Hanbury Brown radiotelescope~\cite{hanbury-brown52nat},
of the Allred radiometer~\cite{allred62jrnbs,arthur64im}, and of the
Johnson/Nyquist thermometry~\cite{white96metro}.  Obviously, the
interferometer fluctuation can not be rejected because there is a
single interferometer shared by the two channels, but it would be if
the interferometer was be duplicated.

\subsection{$\pm45\degrees$ Detection}\label{sec:quadrature-correlation}

In the $\pm45\degrees$ correlation mode only one radiofrequency
channel is used, and the matrix $R$ is set for a frame rotation of
45\degrees, thus equations \req{eqn:mixer:i:out} and
\req{eqn:mixer:q:out} become
\begin{eqnarray}
\displaystyle
w_1(t) & = & \sqrt{\frac{g}{8\ell_m}} \; \left[  
             \frac{1}{\sqrt{2}}\;n_1(t) + \frac{1}{\sqrt{2}}\;n_2(t)
             \right] \\
\displaystyle
w_2(t) & = & \sqrt{\frac{g}{8\ell_m}} \; \left[  
             \frac{1}{\sqrt{2}}\;n_1(t) - \frac{1}{\sqrt{2}}\;n_2(t)
             \right] \;\;,
\end{eqnarray}
and therefore
\begin{equation}
S_{12}(f)= \frac{g}{4\ell_m} \; \left[N_1(f) - N_2(f)\right] \;\;.
\end{equation}
The trick is that with true random noise, including thermal noise,
$n_1(t)$ and $n_2(t)$ have identical statistical properties, hence
$S_{12}(f)=0$.  Conversely, when a random process modulates a
parameter of the DUT, it tends to affect the phase of the carrier and
to let the amplitude unchanged, \emph{or} to affect the amplitude and
to let the phase unchanged.  Obviously, this depends on the physical
phenomena involved, the knowledge of which is needed for the
instrument to be useful.  This type of detection was originally
invented for the measurement of electromigration in metals at low
frequencies~\cite{verbruggen89apa}, which manifests itself as a random
amplitude modulation, and then extended to the measurement of phase
noise of radiofrequency devices~\cite{rubiola00eftf}.

\section{Implementation}

\begin{figure}[t]
\centering\includegraphics[width=\textwidth]{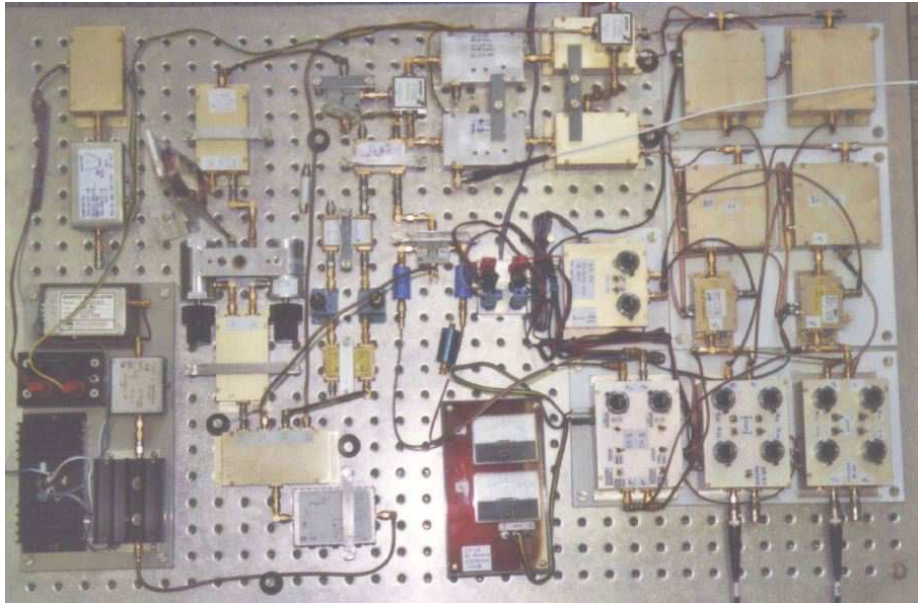}
\caption{Picture of the described prototype.}
\label{fig:photo}
\end{figure}

We constructed a prototype, shown in Fig.~\ref{fig:photo} and
described underneath, designed for the carrier frequency $\nu_c=100$
MHz.

In search of the highest sensitivity at low frequencies, we decided
not to use commercial hybrids or power splitters in the inner
interferometer.  In fact these devices are based on ferrite inductors
and transformers, for they could flicker by modulating the carrier
with the magnetic noise of the core; in addition, they could generate
harmonics of the carrier frequency, noxious to the amplifier
linearity.  Conversely, the two couplers between the inner
interferometer and the amplifiers can be of the ferrite type because
they are crossed by a low residual power, or by the noise sidebands
only.  Thus we built two Wilkinson couplers, each with a pair of
$\lambda/4$ 75 \ohm\ PTFE-insulated cables and a 100 \ohm\ metal
resistor.  After trimming for best isolation at 100 MHz, the
dissipative loss and the isolation turned out to be of 0.15 dB and 34
dB, respectively.

The phase of the inner interferometer can be adjusted by means of a
set of semirigid cables and SMA transitions.  In some experiments we
also used a type of microwave line-stretcher consisting of coaxial
pipes with locknuts, whose internal contacts are well protected
against vibrations; the popular and easy-to-use U-shaped line
stretcher adjusted by means of a micrometer is to be avoided.  The
attenuation can be adjusted with commercial 0.1 dB step attenuators.
Two types were tested, manufactured by Weinschel (mod.\ 3035) and
Texscan (mod.\ MA-508), with almost identical results.  Measuring the
ultimate noise of the instrument, we used only fixed attenuators and
cables, manually trimmed.  The instrument can still be used in this
way for actual applications, provided the attenuator and the phase
shifter be matched to the specific DUT\@.

The 100 MHz amplifiers consists of three cascaded modules in bipolar
technology with low-Q LC filters in between.  The first filter is a
capacitive-coupled double resonator, while the second one is a LCL
T-network.  The use of two different topologies warrants a reasonable
stopband attenuation at any frequency of interest because the stray
pass frequencies do not coincide.  For best linearity, we used for the
second and for the third stage a type of amplifier that shows a third
order intercept point of 35 dBm and a 1 dB compression power of 17
dBm, while the total output power never exceeds some $-50$ dBm.  The
complete amplifier shows a gain $g=41$ dB and a noise figure $F=1.5$
dB\@.

The I-Q modulator and the I-Q detector are two equal devices built for
this purpose (Fig.~\ref{fig:iq}).  The dissipative losses of the power
splitter and of the 90\degrees\ hybrid are of 0.5 and 1 dB
respectively, while the SSB loss of the mixers is $\ell_m=6$ dB\@.  A
10.7 MHz low pass filter is inserted at the IF port of each mixer to
block the $2\nu_0$ image frequency and the $\nu_0$ crosstalk.  The
quadrature adjustment, present in the earlier releases, is no longer
useful because the quadrature error is compensated by the $R$ and $D$
matrices.  The harmonics of the I-Q modulator must be checked; in our
case none of them exceed $-65$ dBm, which still insufficient for the
amplifier to flicker.  I-Qs of similar performances and smaller size,
are available off-the-shelf at a lower cost.  To duplicate the
instrument this is the right choice; we opted for the realization of
our own circuit for better insight, useful in the very first
experiments.  Finally, self interference from the mixer pump signal to
the amplifier can be a serious problem, which caused some carefully
designed layouts not to work properly.

The master source is a high stability 100 MHz quartz oscillator
followed by a power amplifier and by a 7-poles LC filter that removes
the harmonics.  The output power is set to 21 dBm, some 8 dB below the
1 dB compression point.  The source exhibits a frequency flicker of
$-127$ \unit{dBrad^2/Hz} at $f=100$ Hz and a white phase noise of
$-155$ \unit{dBrad^2/Hz}.

The preamplifiers at the detector output are a modified version of the
``super low noise amplifier''~\cite{pmi:mat03spec}, consisting of three PNP
matched differential pairs connected in parallel and followed by a low
noise operational amplifier.  The preamplifier, that shows a gain of
52 dB, is still not optimized\footnote{A version of this amplifier optimized for 50~\ohm
sources was studied afterwards~\cite{rubiola04rsi}} for the input impedance of 50~\ohm.
Terminating the input to 50~\ohm, the overall noise (preamplifier and
termination) is of 1 \unit{nV/\sqrt{Hz}} (white), and 1.5
\unit{nV/\sqrt{Hz}} at 1 Hz (flicker plus white).
The dc offset necessary to compensate for the the
asymmetry of the detector diodes is added at the preamplifier output.

The matrices consist of four 10-turns high quality potentiometers,
buffered at both input and output, and of two summing amplifiers in
inverting configuration.  The coefficients, whose sign is set by a
switch for best accuracy around zero, can be set in the ${\pm}1$
range.

The control consists of two separate integrators based on a FET
operational amplifier in inverting configuration with a pure
capacitance in the feedback path; the capacitors can be discharged
manually.  A dc offset can be added at the output of each integrator,
which is necessary for the manual adjustment of the outer
interferometer to be possible in open loop conditions.  The loop time
constant is of 5 s, hence the cutoff frequency is of 32 mHz.

Each circuit module is enclosed in a separate 4 mm thick aluminum box
with 3 mm caps that provide mechanical stability and shielding.  The
boxes also filter the fluctuations of the environmental temperature.
Microwave UT-141 semirigid cables (3.5 mm, PTFE-insulated) and SMA
connectors are used in the whole radiofrequency section, while high
quality coaxial cables and SMA connectors are used in the baseband
circuits.  All the parts of the instrument are screwed on a standard
$0.6{\times}0.9$ m$^2$ breadboard with M6 holes on a 25 mm pitch grid,
of the type commonly used for optics.  The breadboard is rested on a
500 kg antivibration table that shows a cutoff frequency of 4 Hz.  The
circuit is power-supplied by car-type lead-acid batteries with a
charger connected in parallel; in some cases the charger was removed.

Finally, we implemented a second prototype for the carrier frequency
$\nu_c=5$ MHz.  This instrument is a close copy of the 100 MHz one,
and shares with it the readout system and the carrier control.  Due to
the long wavelength ($\lambda/4=15$ m), the Wilkinson couplers are
impractical.  Provisionally, we use a pair of 180\degrees\ ferrite
hybrids for the inner interferometer.

\section{Adjustment and Calibration}\label{sec:calibration}

For proper operation, the instrument first needs to be tuned and
calibrated.  The first step consists of compensating for the dc offset
due to the diode asymmetry of the I-Q detectors, which is best done
disconnecting the interferometer and terminating the input of the
amplifiers to a 50 \ohm\ resistor.  Secondly, have to set the readout
system, as detailed in Section~\ref{sec:readout}, which also include
the measurement of the SSB gain.  Thirdly, the control loop must be
adjusted according to the procedure given in
Section~\ref{sec:control}, and a suitable time constant must be
chosen.  This turns out to be easier if the inner interferometer is
disconnected and the unused ports are terminated.  Finally, the
interferometer must be set for the highest carrier rejection.

The inner interferometer is first inspected alone with a network
analyzer; as the the phase of the transfer function is not used, a
spectrum analyzer with tracking oscillator is also suitable.  Even at
the first attempt, a slight notch, of at least a fraction of a dB,
appears at some unpredictable frequency.  Hence, the interferometer is
tuned by iteratively `digging' the notch and moving it to the desired
frequency; $\ell$ acts on the carrier rejection, while $\gamma$ acts
on the frequency.  The inner interferometer is then restored in the
machine and the carrier rejection is refined by adjusting the fine
carrier control and inspecting with a spectrum analyzer on the monitor
output of the amplifier.  A rejection of some 80--90 dB should be
easily obtained.  At this stage the machine is ready to use, and a
carrier rejection of 110--120 dB should be obtained in normal
operation.

\subsection{Accuracy}\label{sec:accuracy}

Experience suggests that calibration difficulty resides almost
entirely in the radiofrequency section, while the uncertainty of the
instruments used to measure the low-frequency detected signals is a
minor concern.  In addition, the reference angles are only a second
order problem because an error $\delta\psi$ results in relative error
$-\frac{1}{2}(\delta\psi)^2$ in the measurement of $\phi$ and
$\alpha$, which is negligible in most cases.  The quadrature condition
is even simpler because it is based on a null measurement at the
output of the low-frequency section.

In order to understand calibration, one must remember that 1) $\alpha$
and $\phi$ are voltage ratios, and 2) the instrument circuits are
linear over a wide dynamic range.  As a relevant consequence, the
measurement of $k_\alpha$ and $k_\phi$ relies upon the measurement of
a radiofrequency power ratio instead of on absolute measurements.
Actually, the phase-to-voltage gain (as well as the
$\alpha$-to-voltage gain) is calculated as
$k_\phi=\sqrt{2P_0}\,k\subs{ssb}$, which requires the measurement of
$P_0$.  But the SSB gain is measured with the sideband method and
Equation \req{eqn:k-ssb-meas}.  Consequently,
\begin{equation}
k_\phi=\sqrt{\frac{2P_0}{P_s}}\:W \;\;.
\label{eqn:power-ratio}
\end{equation}
A difficulty arises from the fact that $P_s$ must be a low power,
$-70$ dBm to $-80$ dBm in our case, while $P_0$ can be higher than 10
dBm.  Commercial power meters exhibit accuracy of some 0.1 dB,
provided the input power be not less than some $-30$ dBm; this is
related to the large bandwidth (2--20 GHz) over which the equivalent
input noise in integrated.  Therefore, a reference attenuator is
needed to compare $P_0$ to $P_s$ with a wattmeter.  Actually, we use a
synthesizer followed by a bandpass filter and by a 50 dB calibrated
attenuator to generate the sideband, and we measure the sideband power
at the filter output, before the attenuator; the filter is necessary
to stop the synthesizer spurious signals.  In our case $\nu_0=100$ MHz
is in the frequency range of the two probes (HF-UHF and microwaves) of
the available wattmeter, and we observed that in appropriate
conditions the discrepancy never exceeds 0.05 dB; thus a value of 0.1
dB is a conservative estimate of the wattmeter uncertainty in the
measurement of $P_0/P_s$.  Ascribing an uncertainty of 0.1 dB to the
network analyzer with which the 50 dB attenuator is calibrated, the
estimated accuracy of the instrument is of 0.2 dB.

\section{Experimental Results}\label{sec:results}

Our main interest is the sensitivity of the instrument, that is, the
background noise measured in the absence of the DUT\@.  Obviously, a
great attention is spent on the low frequency part of the spectrum,
where the multiple carrier suppression method is expected to improve
the sensitivity.

In several occasions we have observed that the residual $S_\alpha(f)$
and $S_\phi(f)$ are almost equal, as well as the residual noise
spectrum of any combination $a\alpha+b\phi$ in which $a^2+b^2=1$.
Rotating the detection frame with the matrix $B$, the variation of the
residual flicker can be of 1 dB peak or less.  This means that the
residual noise $N\subs{rf}$ has no or little preference for any angle
versus the carrier.  Hence, after putting right the phase modulator
method we did not spend much effort in calibrating the detection
angle.  Thus, the detected noise is the scalar projection of
$N\subs{rf}$ on two orthogonal axes that in most cases we let
arbitrary.  On the other hand, it would be misleading to give the
results in terms of $N\subs{rf}$ because the parametric noise is
affected by the carrier power, and because the ratio $N\subs{rf}/P_0$
is needed to determine $S_\phi(f)$ and $S_\alpha(f)$.  Therefore, we
give the results in terms of the normalized noise
$S_n(f)=N\subs{rf}/P_0$.  Of course, $S_n(f)$ becomes $S_\phi(f)$ or
$S_\alpha(f)$ if $B$ operates the appropriate rotation.  The unit of
$S_n(f)$ is
\unit{[rad^2]/Hz}, hence \unit{dB[rad^2]/Hz}, where \unit{[rad^2]}
implies that the unit of angle appears in the appropriate conditions.
Anyway, the presence or absence of the unit \unit{rad^2} has no effect
on numerical values.  As in real applications the measured quantities
will be `true' phase and amplitude noise, all the plots are labeled as
$S_\phi(f)$ and $S_\alpha(f)$, given in
\unit{dBrad^2/Hz} and \unit{dB/Hz}.  
Yet, in order to avoid any ambiguity, the radiofrequency spectrum
$N\subs{rf}$ is also reported (in dBm/Hz) and it is always specified
whether or not the angle is calibrated.

Finally, the laboratory in which all the experiments are made is not
climatized; a shielded chamber is not available, therefore the
electromagnetic environment is relatively unclean.  A 100 Mbit/s
computer network is present, while the electromagnetic field of FM
broadcastings in the 88--108 MHz band is of the order of 100
\unit{dB\mu V}.  Even worse, our equipment is located over the top of
a clean room for Si technology where several dreadful (for us)
machines are operated regularly, like vacuum pumps, an elecron
microscope, etching and ion sputtering systems, etc., and we also
share the power-line transformer with the clean room.  No attempt has
been made to hide stray signals by post-processing, consequently all
the reported spectra are true hardware results.

\subsection{Lowest-Noise Configuration}\label{sec:lowest-noise-config}
%
The first set of experiments is intended to assess the ultimate
sensitivity of the instrument.  Therefore the inner interferometer is
balanced with semirigid coaxial cables only.  In this conditions there
results an asymmetry of a fraction of a degree in phase, and of
several hundredth of dB in amplitude, which is corrected by inserting
a parallel capacitance and a parallel resistance in the appropriate
points, determined after some attempts.  The actual correction is so
small---some 0.5 pF and a few k\ohm\ in parallel to a 50 \ohm\
line---that the resulting impedance mismatch has no effect on the
noise measurement accuracy.  In the reported experiment the carrier
rejection is of 88 dB in $\Delta'$.  While the automatic carrier
control is operational, the fine control, no longer needed, is
disconnected.  With a DUT power $P_0=14.1$ dBm, the gain is
$\sqrt{P_0}k\subs{dsb}=80.5$ \unit{dBV[/rad]}\@.

\subsubsection{Single-Arm Mode, for Real-Time Operation}\label{sec:results-single-arm}
%
\begin{figure}[t]
\centering\includegraphics[bb=0 0 242 158, scale=1]{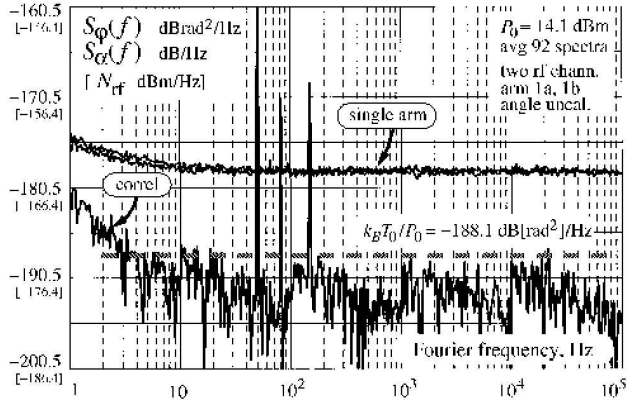}
\caption{Ultimate residual noise, measured in the absence of the DUT
  and with fixed-value devices in the interferometer.}
\label{fig:plot-471}
\end{figure}
Figure~\ref{fig:plot-471} shows the residual noise spectrum at the
output 1 of the two radiofrequency channels and the cross spectrum.
The detection direction, let arbitrary, is the same for the two
channels.  The white noise floor is $N\subs{rf\,0}=-165$ dBm/Hz, which
is 9 dB above the thermal energy $k_BT_0=-174$ dBm/Hz.  This
relatively high value is due to the high loss of the DUT-amplifier
path, which is of 7.5 dB; this includes the 6 dB intrinsic loss of the
couplers CP2 and CP4 and the insertion loss of CP3.  The amplifier
contributes with its noise figure $F=1.5$ dB\@.  The noise floor
corresponds to $S_{n0}=-179.1$ \unit{dB[rad^2]/Hz}.  Of course, if one
radiofrequency channel is removed and the coupler in between (CP4) is
bypassed, the gain $k\subs{dsb}$ increases by 3.5 dB while the white
noise voltage at the output is still the same.  Consequently the noise
floor becomes $S_{n0}=-182.1$ \unit{dB[rad^2]/Hz}.

On the left of Fig.~\ref{fig:plot-471}, at $f=1$ Hz, the residual
noise is of $-161.4$ dBm/Hz (channel $a$) and $-161.0$ dBm/Hz (channel
$b$), which corresponds to a normalized noise $S_n(1\unit{Hz})$ of
$-175.5$ \unit{dB[rad^2]/Hz} and $-175.1$ \unit{dB[rad^2]/Hz}.  After
correcting for the white noise contribution, the true flicker is of
$-178.0$ \unit{dB[rad^2]/Hz} and of $-177.3$ \unit{dB[rad^2]/Hz}, for
the two channels.

\begin{figure}[t]
\centering\includegraphics[bb=0 0 239 155,scale=1]{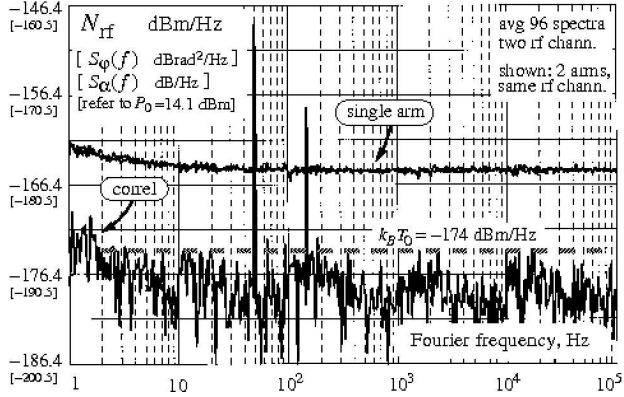}
\caption{Residual noise of the radiofrequency electronic
  circuits, measured in the absence of the interferometer.}
\label{fig:plot-478}
\end{figure}

In a second experiment, the interferometer is removed, and the common
input of the two radiofrequency channels (the input of CP4) is
terminated.  The automatic carrier control, still operational,
compensates only for leakage.  This experiment is intended to divide
the noise of the amplifier and detector from that of the
interferometer.  Figure \ref{fig:plot-478} shows the residual noise of
the two arms of the same radiofrequency channel, accurately set in
quadrature with one another.  Obviously, only $N\subs{rf}$ can be
measured because there is no carrier to normalize to.  Anyway,
$S_{\phi}(f)$ and $S_{\alpha}(f)$ are also reported for comparison,
taking a fictive carrier power of the same value.

Comparing Fig.~\ref{fig:plot-478} to Fig.~\ref{fig:plot-471}, the
noise floor $N\subs{rf\,0}$ is unchanged, which means that the white
noise of the interferometer is negligible.  Without interferometer,
$N\subs{rf}$ is of $-162$ dBm/Hz at $f=1$ Hz, which is some 1 dB lower
than the previous value.  This indicates that most of the flicker of
Fig.~\ref{fig:plot-471} comes from the amplifier and detector, and
that the interferometer noise is some 6--7 dB lower than that appears
from Fig.~\ref{fig:plot-471}, say $-182$ \unit{dB[rad^2]}/Hz at $f=1$
Hz.

\subsubsection{Correlation and Averaging}\label{sec:results-correlation}

Back to Fig.~\ref{fig:plot-471}, the low-frequency correlation between
the two channels is of $-168.2$ dBm/Hz at $f=1$ Hz, hence
$S_{ab}(1\unit{Hz})=-182.3$ \unit{dB[rad^2]/Hz}.  This is the
stability of the interferometer, shared between the two channels.  In
fact, a noise reduction of $\sqrt{2m}=11.3$ dB would be expected if
the two channels were independent, while the actual noise reduction is
only of some 6.5 dB\@.  In addition, this confirms the sensitivity
inferred in Section~\ref{sec:results-single-arm}, when we removed the
interferometer.  Smoothing the plot of Fig.~\ref{fig:plot-471}, the
correlated noise is lower than $k_BT_0/P_0$ at a Fourier frequency as
low as 3 Hz.

\begin{figure}[t]
\centering\includegraphics[bb=0 0 239 159,scale=1]{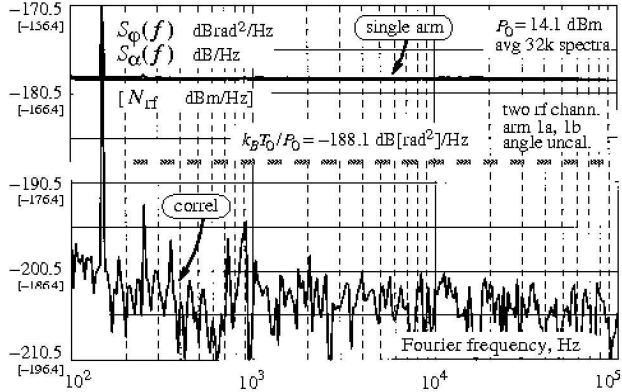}
\caption{Residual noise, measured in the absence of the DUT and
  averaging on a large number of spectra, measured in the same
  condition of Fig.~\protect\ref{fig:plot-471}.  To speed up the the
  experiment frequency spans from 100 Hz.}
\label{fig:plot-473}
\end{figure}

As explained in Section~\ref{sec:parallel-correlation}, the white
noise floor, due to the resistive terminations and to the amplifiers,
is expected to be rejected in the correlation between the two
channels.  Figure \ref{fig:plot-473} reports the cross spectrum
averaged over $m=32767$ measurements, that is the maximum averaging
capability of the available FFT analyzer.  The observed noise
reduction is close to the value of 24 dB, that is $\sqrt{2m}$.
Therefore, there is no evidence of correlated noise, and the
sensitivity is expected to further increase increasing $m$.  In the
reported conditions the background noise is $S_{n0}=-203$
\unit{dB[rad^2]/Hz} at $f\ge2500$ Hz, which is 15 dB lower than
$k_BT_0/P_0$.

\begin{figure}[t]
  \centering\includegraphics[bb=0 0 239 172,scale=1]{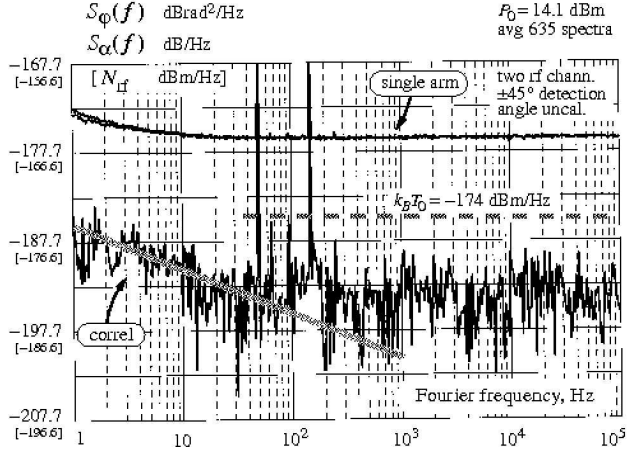}
\caption{Residual noise, measured in the absence of the DUT,
  detecting at ${\pm}45\degrees$ from an arbitrary reference angle.}
\label{fig:plot-474}
\end{figure}

Figure~\ref{fig:plot-474} shows the residual noise spectrum measured
with the two arms of a single radiofrequency channel carefully set in
quadrature with one another, but still referred to an arbitrary
detection direction.  This simulates the detection of phase noise with
the ${\pm}45\degrees$ method.  In the same conditions of the previous
experiments the gain is of 77.8 \unit{dBV[/rad]}, which is 3 dB lower.
This is inherent in the ${\pm}45\degrees$ detection scheme.  As
measurements spanning from 1 Hz take a long time, the experiment was
stopped at $m=635$, well before the cross spectrum could reach its
final value, for $1/\sqrt{2m}=15.5$ dB\@.  The residual noise is
limited by $m$ for $f>10$ Hz, but not in the 1--10 Hz decade.  Fitting
this decade to the $1/f$ slope results in a correlated flicker noise
is of some $-186$ \unit{dB[rad^2]/Hz} at $f=1$ Hz, which is the lowest
value we have ever observed.

\subsection{By-Step Attenuator Configuration}\label{sec:by-step-atten}
%
\begin{figure}[t]
  \centering\includegraphics[bb=0 0 239 155,scale=1]{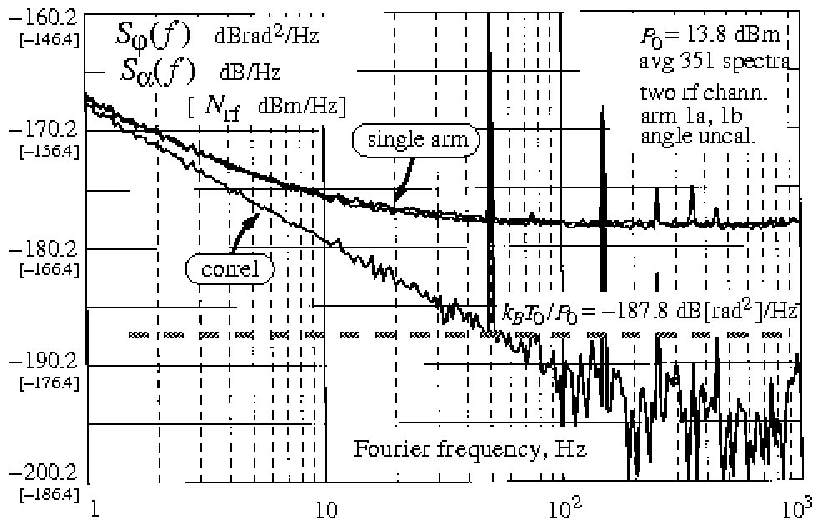}
\caption{Residual noise, measured in the absence of the DUT\@.
  Two by-step attenuators and two microwave coaxial phase shifters
  are present in the inner interferometer.}
\label{fig:plot-459}
\end{figure}
In an easier-to-use version of the instrument, we inserted a by-step
attenuator (Weinschel, mod.\ 3055) and a microwave coaxial phase
shifter in each arm of the inner interferometer, and we restored the
fine carrier control. The phase shifters were set for the best carrier
suppression, while one of the attenuator was set 0.1 dB off the
optimum value, so that the carrier rejection of the inner
interferometer was of 39 dB.  This is slightly worse than the ``true''
worst case, in which a half-step attenuation error of 0.05 dB and a
similar error of the phase shifter result in a carrier rejection of 42
dB\@.  The residual flicker noise of the instrument, shown in the left
part of Fig.~\ref{fig:plot-459}, is of $-168$ \unit{dB[rad^2]/Hz} at
$f=1$ Hz.

Then, we made two additional experiments.  Firstly, we set the
attenuators for the best carrier rejection, and we observed that the
flicker noise does not change.  This means that the 0.1 dB error of
the attenuator is recovered by the fine carrier control without adding
noise, and that the small signal delivered by the closed-loop carrier
control does not impair low frequency sensitivity.  Secondly, we
checked upon the phase shifters with the methods of
Section~\ref{sec:lowest-noise-config}, and we observed a noise
contribution negligible at that level.  The relevant conclusion is
that the flicker noise of Fig.~\ref{fig:plot-459} is due to the
by-step attenuators.  Assuming that the attenuators are equal, each
one shows a flicker noise of $-171$ \unit{dB[rad^2]/Hz} at $f=1$ Hz.
As only one attenuator is needed to measure an actual DUT, this is
also the sensitivity of the instrument.

\subsection{Simplified Configuration}

A simplified version of the instrument is possible, in which the inner
interferometer can be adjusted by step and the fine carrier control is
absent.  Of course, the dynamic range of the closed-loop control must
be increased for the control to be able to recover a half-step error
of the inner interferometer.  This results in higher noise from the
control and in additional difficulty to obtain a slow response.
Actually, this configuration is the first one we experimented on.
\begin{figure}[t]
  \centering\includegraphics[bb=0 0 240 317,scale=1]{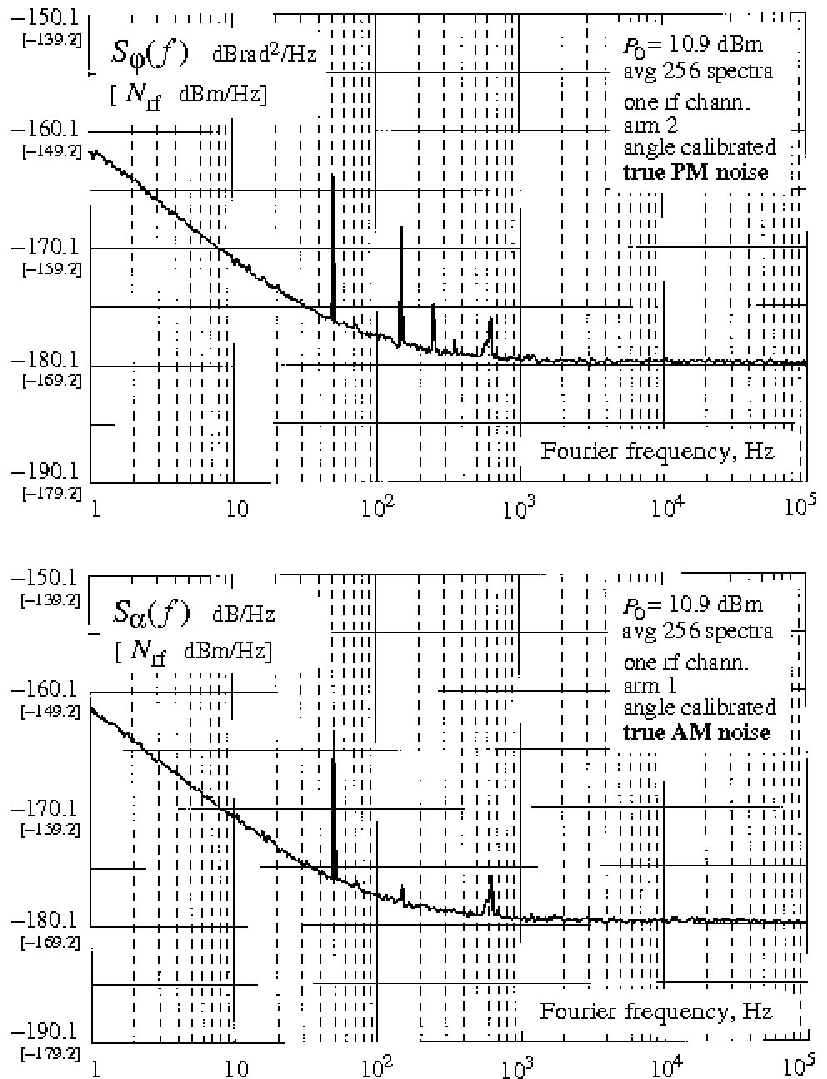}
\caption{Residual noise of the simplified instrument,
  without the fine carrier control.  The detection angle is carefully
  calibrated, for the plots represent true amplitude and phase noise.}
\label{fig:plot-431-430}
\end{figure}

The prototype makes use of two 180\degrees\ hybrid couplers based on
ferrite transformers in the inner interferometer, and has only one
radiofrequency channel.  Operating at $P_0=10.9$ dBm, the gain is of
80.1 \unit{dBV[/rad]}.  The direction of detection was calibrated
carefully, therefore in this case the residual noise consists of true
phase noise and of true amplitude noise.  In order to simulate the
worst case, we first trimmed the inner interferometer for a relatively
deep minimum of the residual carrier, and then we set the attenuator
0.1 dB off that point.  The residual noise spectra are shown in
Fig.~\ref{fig:plot-431-430}.  The white noise is $S_{\alpha0}=-179.6$
\unit{dB/Hz} and $S_{\phi0}=-179.6$ \unit{dBrad^2/Hz}.  This is equal
to the expected value $2Fk_BT_0/P_0\ell$, where $\ell=0.8$ dB accounts
for the dissipative loss in the DUT-amplifier path and for the
insertion loss of the 20 dB coupler; the noise figure of the amplifier
is $F=1.5$ dB\@.  The residual flicker is
$S_\alpha(1\unit{Hz})=-161.5$ \unit{dB/Hz} and
$S_\phi(1\unit{Hz})=-161.4$ \unit{dBrad^2/Hz}, which is ascribed to
the closed-loop carrier control.

\section{Measurement Examples}

The main conclusion of Section~\ref{sec:by-step-atten}, that the
flicker noise of a by-step atteuator is of $-171$ \unit{dB[rad^2]/Hz}
at $f=1$ Hz, is a first example of measurement out of reach for other
instruments.

\begin{figure}[t]
\centering\includegraphics[scale=1]{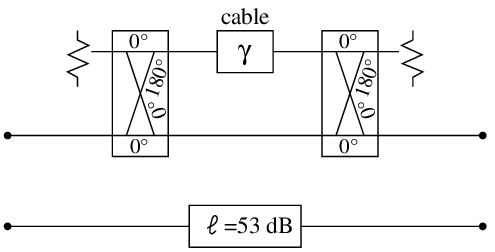}
\caption{Measurement of the noise of a pair of HH-109 hybrid couplers,
  inserting the hybrid pair as the inner interferometer.}
\label{fig:hybrids-meas-scheme}
\end{figure}

\begin{figure}[t]
\centering\includegraphics[bb=0 0 239 155,scale=1]{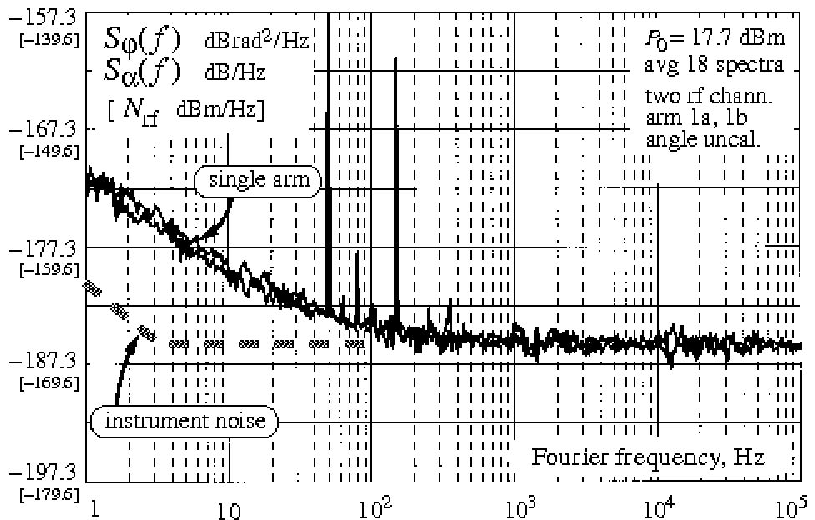}
\caption{Noise of a pair of HH-109 hybrid couplers, measured 
  at $\nu_0=100$ MHz.  Assuming that the devices are equal, the noise
  of each is 3 dB lower than shown.}
\label{fig:plot-469}
\end{figure}

As a second example of application, we measured a pair of 180\degrees\ 
hybrid couplers (HH-109 Anzac, now Macom) with the scheme of
Fig.~\ref{fig:hybrids-meas-scheme} (top) inserted as the inner
interferometer.  The difference between the two samples turned out to
be so small that a carrier rejection of 53 dB could be achieved by
exploring the combinatorial permutations of the geometrical
configuration.  Hence, the background noise of the instrument was
tested by replacing the hybrid pair with a 53 dB attenuator
(Fig.~\ref{fig:hybrids-meas-scheme} bottom).  As the device noise
detected on two orthogonal axes was almost the same, we did not
calibrate the detection angle.  Neglecting losses, the power crossing
the two hybrids is the same because all the input power, 17.7 dBm in
our case, reaches the 50 \ohm\ termination of the second hybrid.
Although we did not use the correlation feature, we did not disconnect
the unused channel.  The result is shown in Fig~\ref{fig:plot-469}.
The noise of the pair is of $-171$ \unit{dB[rad^2]/Hz} at $f=1$ Hz,
while the background noise is of $-180.5$ \unit{dB[rad^2]/Hz}.  After
subtracting the latter, the flicker noise of the pair is of $-171.5$
\unit{dB[rad^2]/Hz}, and therefore $-174.5$ \unit{dB[rad^2]/Hz} for
each hybrid.

The same HH-109 hybrids, that are designed for the frequency range of
5--200 MHz, were tested at the input power of 14.9 dBm with the 5 MHz
instrument.  In this case we used only one radiofrequency channel, and
we disconnected the other one bypassing the coupler in between (CP4);
this results in a sensitivity enhancement of some 3.3 dB, that
compensates for the reduced driving power.  The measured noise is of
$-171.9$ \unit{dB[rad^2]/Hz} at $f=1$ Hz, thus the flicker noise of
the pair is of $-172.3$ \unit{dB[rad^2]/Hz}, corrected for the the
instrument noise.  Accordingly, the flicker noise of each hybrid is of
$-175.3$ \unit{dB[rad^2]/Hz} at $f=1$ Hz.

The above results confirm the usefulness of the coaxial power dividers
to obtain the highest sensitivity.  As the hybrid is a transformer
network, there are good reasons to ascribe the observed flickering to
the ferrite core.

\section{More about Stability and Residual Noise}
%
\begin{figure}[t]
\centering\includegraphics[scale=1]{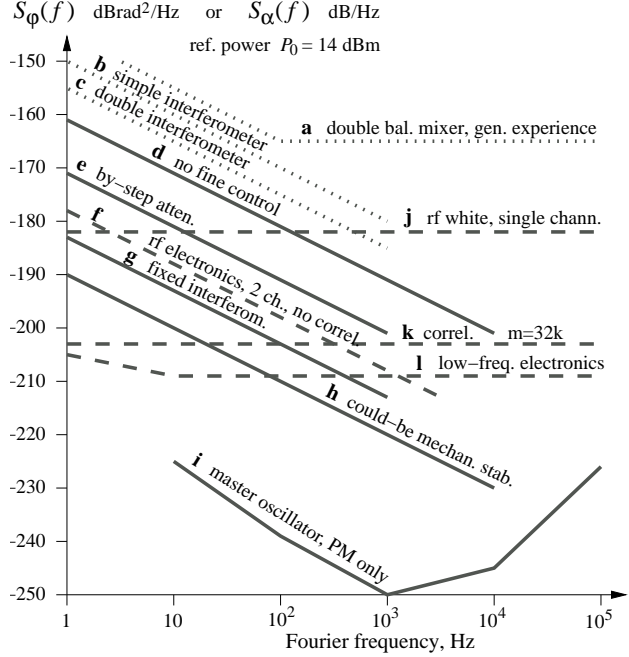}
\caption{Residual noise of the instrument, for different
  configurations, and noise contribution of the most relevant parts.
  For comparison, the dotted lines refer to previous instruments.}
\label{fig:noise-contribs}
\end{figure}
Figure~\ref{fig:noise-contribs} shows a summary of the factors
limiting the instrument sensitivity, most of which taken from
Section~\ref{sec:results}.  For comparison, the dotted lines report
the residual noise of previous instruments: plot a is the double
balanced mixer in average-favorable conditions, while plots b and c
come from our previous works~\cite{rubiola99rsi,rubiola00rsi}.

The limits of the radiofrequency electronics are taken from
Section~\ref{sec:results-single-arm}.  Plot j refers to white noise,
while plot f refers to flickering corrected for white.  The
correlation limit (plot k) is the white noise lowered by
$1/\sqrt{2m}$.  The noise of the baseband electronics (plot l) is
measured terminating the preamplifier input to 50 \ohm, and referring
the output noise voltage $S_{v0}$ to the DUT\@.  Plots f, j, k and l
(dashed lines) are related to $N\subs{rf}$ levels independent of
$P_0$, for the corresponding $S_\phi(f)$ and $S_\alpha(f)$ decrease as
$P_0$ increases.  A conventional power $P_0=14$ dBm is assumed.

Noise from the master oscillator (plot i) is measured with a phase
modulator between the oscillator and the power amplifier, scaling the
result down according to the actual oscillator noise.  The modulation
needed is of some 80--100 dB higher than the oscillator noise, for
this only proves that the oscillator phase noise is negligible,
without providing a precise result.  Unfortunately, we have no
information about the amplitude noise of the oscillator.

The ``could-be mechanical stability'' (plot h) is a reference value
inferred from the residual flicker of $-162$ dB at $f=10$ Hz that we
measured measured at 9.1 GHz on our first interferometer~\cite{rubiola99rsi},
under the obvious assumption that the mechanical fluctuations could
not be worse than the overall noise we measured.  We guess that the
above result can be scaled down by 39 dB, which is the ratio
(9.1\,GHz)/(100\,MHz), assuming a similar fluctuation in length.  A
phase fluctuation $\phi(t)$ is equivalent to a length fluctuation
$l(t)=(\lambda_c/2\pi)\phi(t)$, where $\lambda_c\simeq2.4$ m is the
wavelength inside cables.  Hence, the value of $-182$
\unit{dBrad^2/Hz}, taken as a conservative estimate of the
interferometer flicker (Sections~\ref{sec:results-single-arm} and
\ref{sec:results-correlation}), is equivalent to
$S_l(1\unit{Hz})=9.2{\times}10^{-20}$ \unit{m^2/Hz}.  In the case of
flicker noise, the appropriate formula to convert the PSD $S_y(f)$ of
the quantity $y$ into the Allan variance $\sigma^2_y(\tau)$ is
$\sigma^2_y(\tau)=2\ln2\,S_y(1\unit{Hz})$, independent of the
measurement time $\tau$, as well known in the domain of time and
frequency metrology~\cite{rutman78pieee}.  In our case the Allan deviation,
that is the stability of the interferometer, is $\sigma_l=3.6$ \AA\@.
The latter is far from the stability achieved by other scientific
instruments, like the scanning microscope, for we believe that there
is room for progress.

\paragraph{Acknowledgements.}
Michele Elia, from the Politecnico di Torino, Italy, gave invaluable
contributions to the theoretical comprehension of our experiments.
Hermann Stoll, from the Max Planck Institute for Metallforschung,
Stuttgart, taught us the trick of the ${\pm}45\degrees$ detection.
Giorgio Brida, from the IEN, Italy, shared his experience on
radiofrequency.  The technical staff of the LPMO helped us in the
hardware construction.

\def\bibfile#1{/home/rubiola/docs/bib/#1}
\bibliographystyle{amsalpha}
\bibliography{\bibfile{ref-short},%
              \bibfile{references},%
              \bibfile{rubiola}}

\end{document}